\documentclass[12pt,a4paper]{panl}
\usepackage{cite}
\usepackage{wrapfig}
\usepackage{graphicx}
\usepackage{amssymb}
\usepackage{amsfonts}
\usepackage{amsmath}
\usepackage{longtable}
\usepackage{rotating}
\usepackage{lscape}
\usepackage{epsfig}
\usepackage{multirow}

\originalTeX
\begin{document}
% Journal sections (see http://pkp.jinr.ru/index.php/PEPAN_LETTERS/about/editorialPolicies#focusAndScope)
%\issuearea{Physics of Elementary Particles and Atomic Nuclei. Theory}
% or in Russian
%\issuearea{ФИЗИКА ЭЛЕМЕНТАРНЫХ ЧАСТИЦ И АТОМНОГО ЯДРА. ТЕОРИЯ}

\title{ Exclusive $J/\Psi$ Production and Gluon GPDs Effects}
\maketitle
\authors{S.V.\,Goloskokov$^{a,}$\footnote{E-mail:  goloskkv@theor.jinr.ru},
Ya-Ping\,Xie$^{b,}$\footnote{E-mail: xieyaping@impcas.ac.cn}}
\setcounter{footnote}{0}
\from{$^{a}$\,BLTP, Joint Institute for Nuclear Research, Dubna 141980, Moscow region, Russia}
\from{$^{b}$\,Institute of Modern Physics, Chinese Academy of Sciences, Lanzhou 730000, China}

\begin{abstract}
Study of  exclusive photoproduction of $J/\Psi$ mesons was carried out in
a factorization approach. Generalized parton distributions (GPDs) for gluons,
which play an important role here, are constructed using the double distribution
representation. The obtained cross sections of $J/\Psi$ production in a wide energy
range are in good agreement with the experiment. The spin asymmetries in the
$J/\Psi$ production were estimated. These calculations were extended to the of
$J/\Psi$ mesons production in ultraperipherical $pp$ collisions. It was shown a
good agreement of our results with the LHCb data. Predictions have been made for
the NICA energy range. The  model results can be used to study the gluon GPDs
in future colliders.
\end{abstract}
\vspace*{6pt}


\label{sec:intro}
\section*{Introduction}
In this report, we  analyze  exclusive production of $J/\Psi$ mesons which can
be used to study gluon GPDs. GPDs are important objects that contain extensive
information about the hadron structure. GPDs depend on three variables:
longitudinal momentum fraction $x$, skewness $\xi$- longitudinal momentum transfer
fraction, and momentum transfer, $t$. In the forward limit, GPDs are equal to
parton distribution functions (PDFs). The GPDs moments are connected with hadron
form factors and contain information on the parton angular momenta. This means
that GPDs give information on the 3D  structure of hadrons (see e.g. review
\cite{GPD-review01, GPD-review02, burk} and references there).

An important processes to study GPDs are deeply virtual Compton scattering,  light  meson leptoproduction at
high photon virtuality and some other reactions where  the gluon and quark GPDs can be analyzed, see e.g. \cite{dvcs,dvmp2}. The amplitudes of  these reactions  factorize into the hard subprocess and  GPDs \cite{dvcs,ji,rad}. It is well known that heavy meson production gives information on gluon GPDs in proton \cite{rys1,rys2}.

This report is devoted to studying the production of heavy $J/\Psi$ mesons in lepton-proton and proton-proton reactions.
In section 2, we present the handbag approach where the exclusive $J/\Psi$ meson photoproduction amplitudes
can be presented as a convolution of hard subprocess  and soft GPDs \cite{jpsi}. The gluon GPDs $H_g$ are constructed using double distribution representation.

In  section 3, we show a comparison of model results for the $J/\Psi$ production cross section with experimental data from HERA to LHC energies \cite{jpsi} which described fine. We estimate $J/\Psi$ spin asymmetries
at the energies of future electron-ion colliders in USA and China \cite{eic,eicc} where information on $E_g$ and $\tilde H_g$ GPDs can be obtained.

In  section 4, we extend our model calculations to exclusive $J/\Psi$ production  in
proton-proton collisions \cite{jpsipp}. The survival factors and equivalent
photon approximation are applied to predict the exclusive
$J/\Psi$ production in pp collisions. Obtained results are in a good
agreement with the experimental data at LHCb. We predict the exclusive
$J/\Psi$ production cross section at NICA energies.

Our GPD approach predictions can be employed to estimate the
exclusive $J/\Psi$ production observables in future lepton–proton and proton–proton experiments
to get essential information on gluon GPDs.

\section*{Handbag approach for heavy  meson leptoproduction}

The meson production amplitude $\mathcal{M}$ in Eq.~(3) can be presented in the form \cite{gk06,jpsi}
\begin{eqnarray}
\mathcal{M}_{\mu^\prime+,\mu+}
&=& \frac{e}{2}C_V\int_0^1\frac{dx}{(x+\xi)(x-\xi+i\epsilon)}\mathcal{H}_{\mu^\prime,\mu}\,H_g(x, \xi, t),\\ \nonumber
\mathcal{M}_{\mu^\prime-,\mu+}
&=&-\frac{e}{2}C_V\frac{\sqrt{-t}}{2m}\int_0^1\frac{dx}{(x+\xi)(x-\xi+i\epsilon)}\mathcal{H}_{\mu^\prime,\mu}\,E_g(x, \xi, t).
\end{eqnarray}
Here the proton non-flip  and the helicity flip amplitudes contain the hard part $\mathcal{H}$ that is calculated perturbatively and soft gluon GPDs $H_g$ and $E_g$, respectively.

We consider 6 gluon diagrams to calculate hard scattering amplitude \cite{gk06,jpsi}. The hard amplitude $\mathcal{H}$  can be written as
\begin{equation}\label{hsaml}
  {\cal H}_{\mu'+,\mu +}\,\simeq
\,\alpha_s(\mu_R)
            \,\int_0^1 d\tau\,\int \frac{d^{\,2} k_\perp}{16\pi^3}
            \phi_{V}(\tau,k^2_\perp)\;
                f_{\mu',\mu}(x,\xi,\tau) \hat{D} (Q^2,k^2_\perp).
\end{equation}
In calculations we take into account the transverse part of quark momenta $k_\perp$ in the soft meson wave function $\phi$ and in quark propagators, which are collected in the $D$ function
\begin{eqnarray}
D=
\frac{1}{(2\mathbf{k}^2_\perp+\mu_V^2)(4\xi\mathbf{k}_\perp^2+\mu_V^2(\xi-x)+i\epsilon)(4\xi\mathbf{k}_\perp^2+\mu_V^2(\xi+x))}.
\end{eqnarray}
This form is valid for the case of the nonrelativistic  wave function $\phi_{V}$ in Eq.~(2) with the same quark and antiquark momenta ($\tau=1/2$).  The average value of $k_\perp^2$ in propagators Eq.~(3) is determined by the wave function. We find that  $<k_\perp^2>\sim 1\mbox{GeV}^{2}$  describes experimental data well.

A typical variable (scale) in the hard propagators $D$ is $\mu_V^2=m_V^2+Q^2$   which is large for $J/\Psi$ production that provides the amplitude factorization even for $Q^2=0$. The hard amplitudes $f$ in (2) for longitudinally and transversally polarized photon and vector meson have a form
\begin{eqnarray}
f_{11}=64\sqrt{Q^2}\mu_V^4(x^2-\xi^2),\;\;\;f_{00}=-\frac{\sqrt{Q^2}}{m_V}\,f_{11}.
\end{eqnarray}
This means that for the same form of wave function for the longitudinally and transversally polarized meson, we have a relation
 \begin{equation}
\mathcal{M}_{0\nu',0\nu}=-\frac{\sqrt{Q^2}}{m_V} \mathcal{M}_{+\nu',+\nu}.
 \end{equation}

To construct GPDs, we use the double distribution representation \cite{mus99} which connects GPDs with PDFs and determines the skewness $\xi$ dependencies of GPDs
\begin{equation}
F_i(x, \xi, t) = \int_{-1}^1d\rho \int_{-1+|\rho|}^{1-|\rho|}d\lambda \, \delta(\rho + \xi \lambda -x)g_i(\rho, \lambda, t).
\end{equation}
where $F_i = H$, $\tilde{H}$, $E$, $\tilde{E}$ GPDs. The double distribution functions $g_i$ for gluon contribution have a form
\begin{eqnarray}
g_i(\rho, \lambda, t) = e^{B_V t}\,h_i(\rho,\mu_F) \frac{15}{16}\frac{[(1-|\rho|)^2-\lambda^2]^2}{(1-|\rho|)^{5}}.
\end{eqnarray}
Here $h_i(\rho,\mu_F)$ is a corresponding $h$ and $e$ gluon PDFs  at the scale $\mu_F \sim \mu_V$, $B_V$ is a slope parameter. Its form can be found in \cite{jpsi}.
We use in our calculations 3  different GPDs models based on PDFs (ABMP, NNPDF, CT). Note, that all references on gluon PDFs and
experimental data on $J/\Psi$ production can be found in \cite{jpsi, jpsipp}.

\section*{$J/\Psi$ production in lepton-proton reactions}

 The differential cross sections  and their transversal part of heavy vector meson in photon-proton scattering are calculated as
\begin{eqnarray}
\frac{d\sigma}{dt}= \frac{d\sigma_T}{dt} (1+ \frac{Q^2}{m_V^2}),\;\;\;\frac{d\sigma_T}{dt}=\frac{1}{16\pi W^2(W^2+Q^2)}
\sum_{\lambda=\pm}\,|\mathcal{M}_{+\lambda,++}|^2.
\end{eqnarray}
The total cross section $\sigma$ is $\frac{d\sigma}{dt}$ integrated over $t$.
\begin{figure}[!h]
	\centering
	\includegraphics[width=6.7cm]{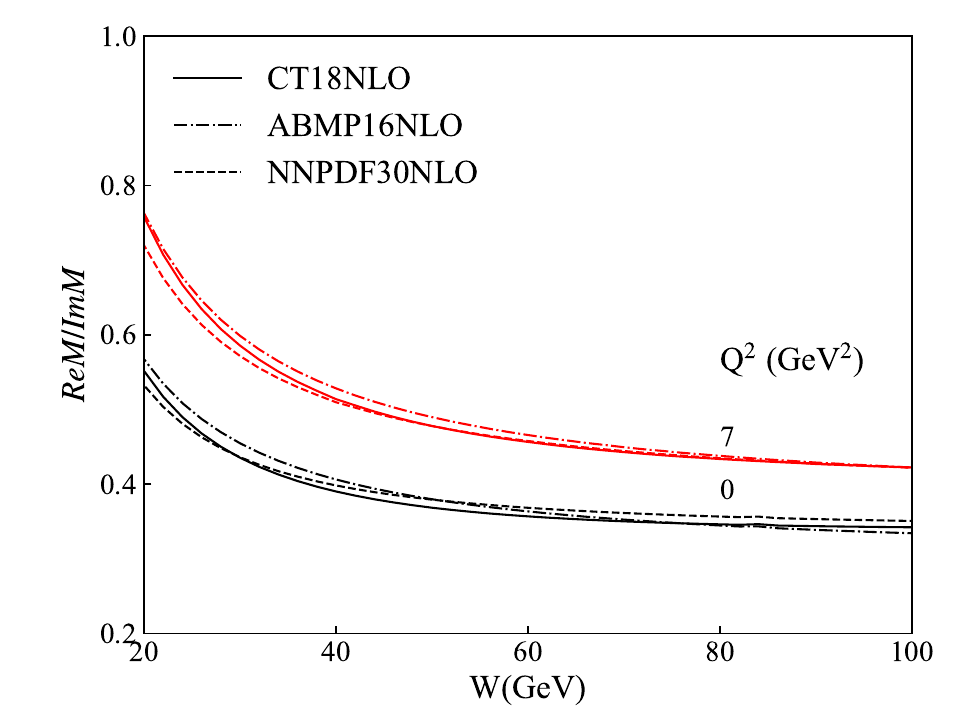}
	\includegraphics[width=6.7cm]{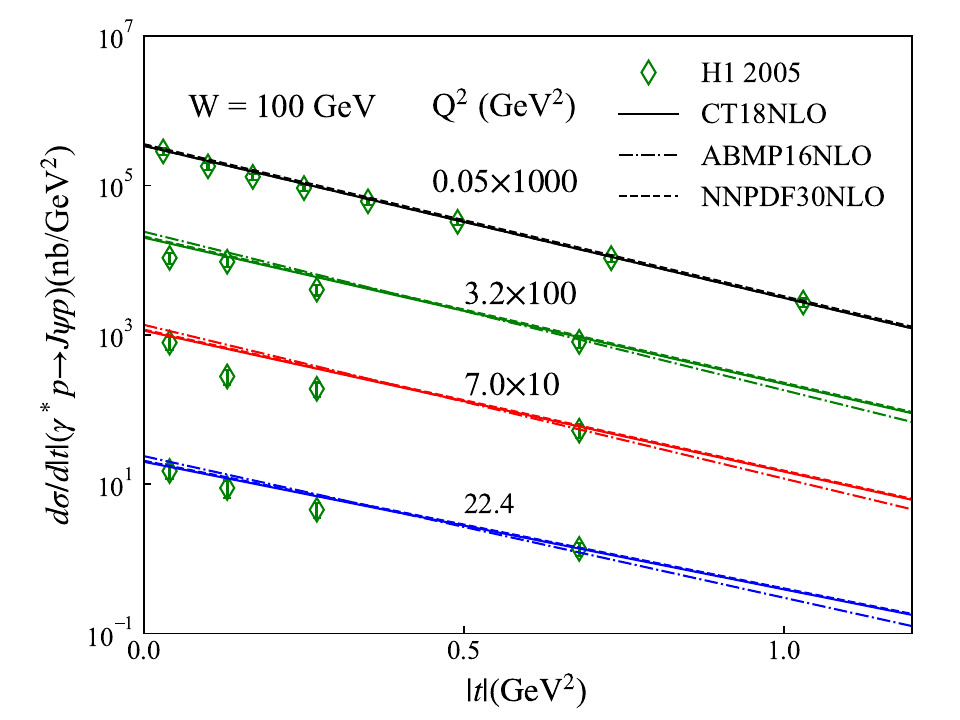}
	\caption{Left:Ratio of $ReM/Im M$ parts of $J/\Psi$ production amplitudes.
		  $J/\psi$  differential cross section vs $|t|$ at  different $Q^2$, right graph. Cross sections are scaled by the factor shown in the graph. }
	\label{fig00}
\end{figure}

\begin{figure}[!h]
	\centering
	\includegraphics[width=6.7cm]{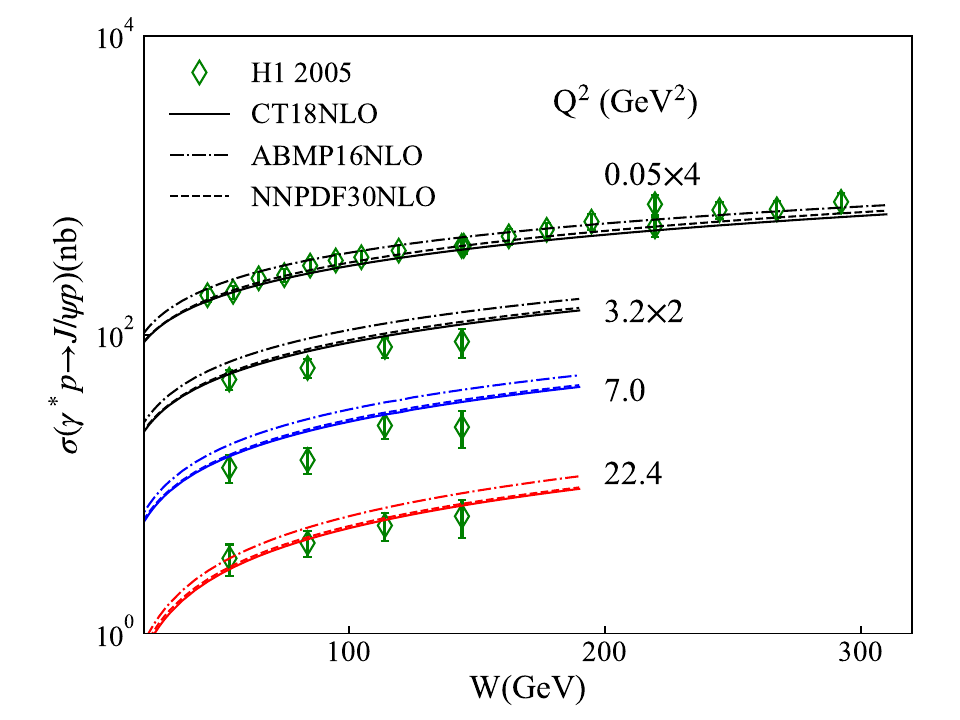}
	\includegraphics[width=6.7cm]{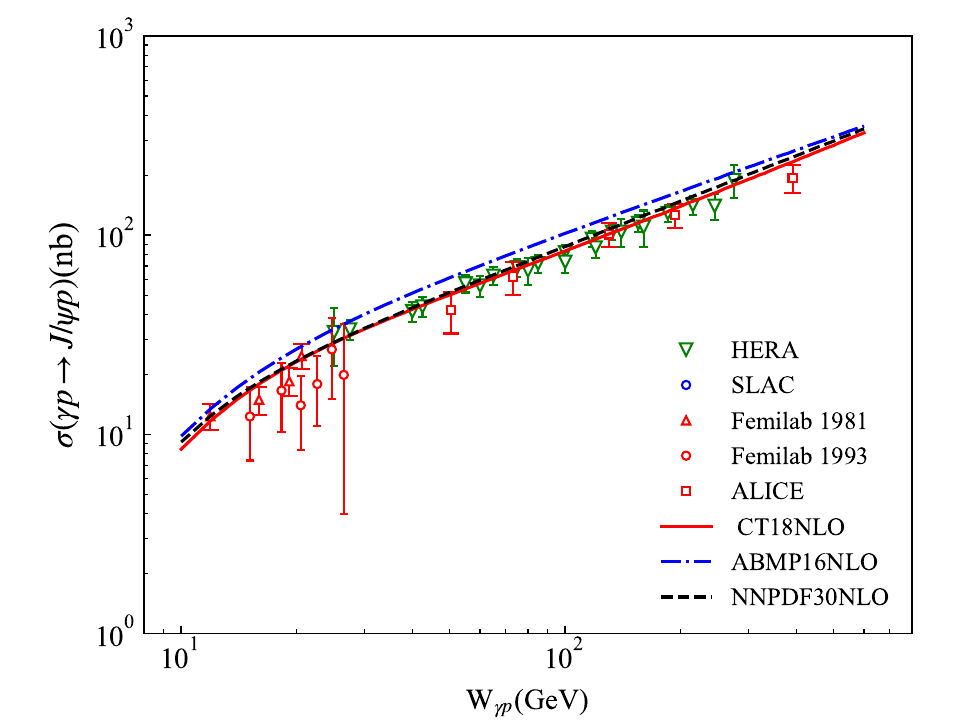}
	\caption{Left: $J/\psi$ total cross section at HERA for fixed $Q^2$
		  vs the energy $W$ and at $Q^2 = 0$ from low to very high energies, right graph. }
	\label{fig01}
\end{figure}
The model gives the possibility to calculate real and imaginary parts of the production amplitude. Our results for the $R=ReM/Im M$ ratio are presented in
 Fig.1, left. R-ratio at fixed $Q^2$ decreases with energy, and at $W=20\mbox{GeV}$ (China EicC) $R(20 GeV) \sim 1$ , at $W=100 \mbox{GeV}$ (USA EiC) $R(100 GeV) \sim 0.5$.

 We compare model results  with available experimental data on the differential cross sections of $J/\psi$ production for H1  at HERA. Our results for differential cross section are in good agreement with data and are shown in Fig. 1, right.
 Results for the total cross section at HERA energies are shown in Fig.2, left, and describe experimental data fine.
  In Fig 2, right the low energy  photoproduction data  are shown together with very high energies data from  LHC ALICE experiment. It can be seen that all GPDs models give good descriptions of experimental data in a wide energy range. More comparison with experiment can be found in \cite{jpsi}

The single spin asymmetry $A_N$ and double spin $A_{LL}$ correlation can  also be measured at heavy vector meson production. They are connected with $E_g$ and polarized $\tilde H_g$ GPDs for gluons correspondingly:
\begin{eqnarray}
A_N
=-\frac{2\mathrm{Im}[\mathcal{M}^H_{++,++}\mathcal{M}^{E*}_{+-,++}]}{|\mathcal{M}^H_{++,++}|^2+|\mathcal{M}^E_{+-,++}|^2},
\;\;\;\;
A_{LL} =
2 \frac{\mathrm{Re}[\mathcal{M}^H_{++,++}\mathcal{M}^{\tilde{H}*}_{++,++}]}
{\epsilon|\mathcal{M}^H_{0+,0+}|^2+|\mathcal{M}^H_{++,++}|^2}
\end{eqnarray}

The $E_g$ GPDs are determine proton spin-flip amplitude Eq.~(1). Previous analyses $A_N$ asymmetry in $J/\Psi$ production can be found in \cite{Koempel:2011rc}.

\begin{figure}[!h]
	\centering
	\includegraphics[width=6.7cm]{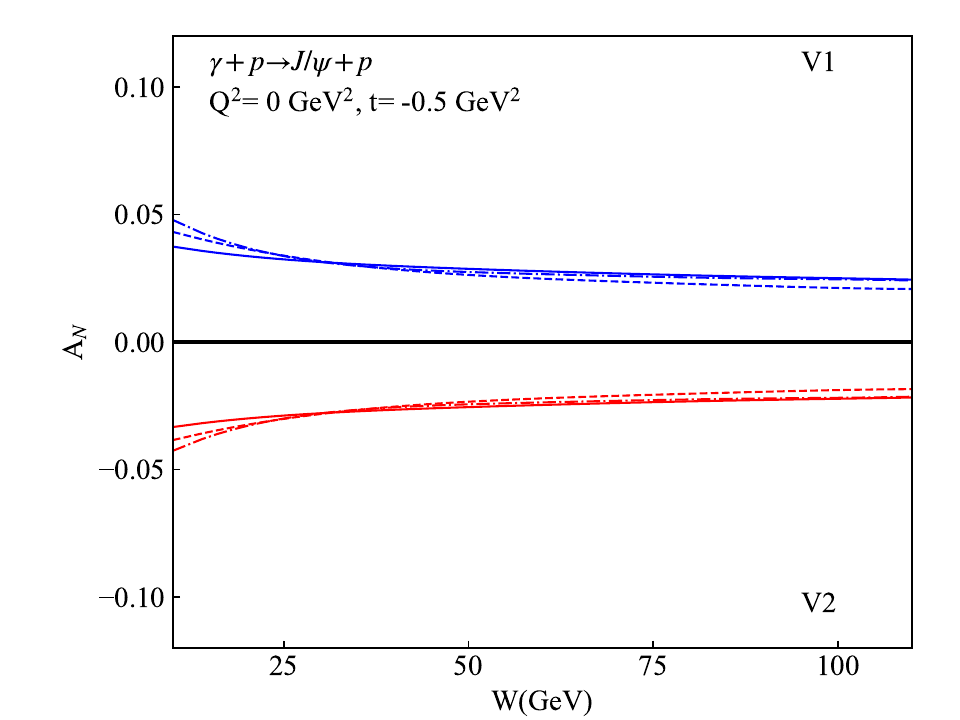}
	\includegraphics[width=6.7cm]{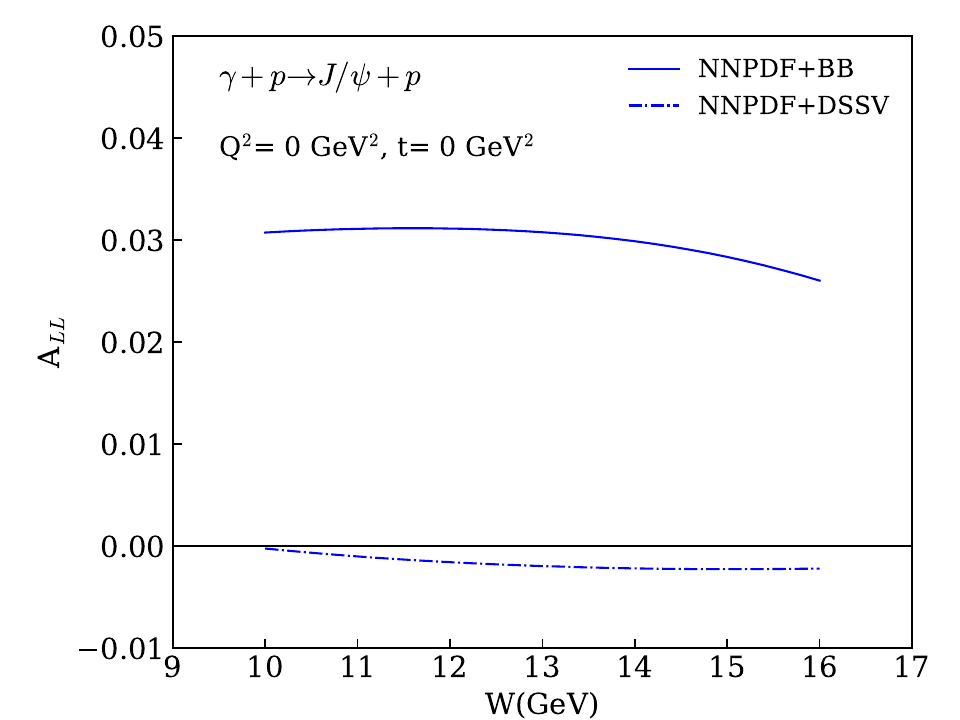}
	\caption{Left: Single spin asymmetry $A_N$ versus W at $Q^2$ = 0 GeV$^2$ and double spin correlation $A_{LL}$, right graph.}
	\label{fig02}
\end{figure}

 To estimate $E_g$ we use parameterizations (2,3) from \cite{gk08}, that called here
  as (V1,V2). They are different mainly by the sign. Energy dependencies of this asymmetry
  is rather  weak because the low -x dependencies of $e_g(x)$ and $h_g(x)$ PDFs are similar.
  Our results on $A_N$  asymmetry for $J/\Psi$ photoproduction are shown in
 Fig.~\ref{fig02}, left. We find that  $|A_N| \sim 0.05$.

Effects of $\tilde H_g$ can be tested in $A_{LL}$ asymmetry.
Details of calculations can be found in \cite{jpsi}.
To study $A_{LL}$ asymmetry, we use BB  and DSSV
parameterizations of $\Delta g$ PDFs which determine $\tilde H^g$
GPDs through DD representation (6). $A_{LL}$ asymmetry decreases rapidly with energy, and at US Eic energies it is close to zero. We show in Fig.\ref{fig02}, right, the $W$ dependencies of  $J/\Psi$ $A_{LL}$ correlation at energies typical for China EicC. We find that $A_{LL} \sim 0.03$ for BB case and
$A_{LL} \sim 0.0$ for DSSV parameterizations.
 Thus, study of $A_{LL}$ correlation at the future EicC can give an answer about the value of $\tilde H^g$.

\section*{$J/\Psi$ production in proton-proton ultraperipheral collisions}

 We adopt our model for $J/\Psi$ production in $pp$ reactions.
 At the proton-proton ultraperipheral collisions, the vector mesons rapidity distributions can be written as follows  \cite{Jones:2016icr}

 \begin{eqnarray}
 \frac{d\sigma(pp)}{dy} = {S^2(W_{+})}{\Big(k_+\frac{dn}{dk_+}\Big)}{\sigma^{J/\Psi}_+(\gamma
 p)}
  +{S^2(W_{-})}{\Big(k_-\frac{dn}{dk_-}\Big)}{\sigma^{J/\Psi}_-(\gamma
  p)}.
 \end{eqnarray}
The photon momentum, $k_{\pm} =x_\gamma ^\pm \sqrt{s_{NN}}/2\approx (m_V/2)e^{\pm |\mathrm{y}|}$. $\mathrm{y}$ is the rapidity of the vector meson, $s_{NN}$ is the collision energy, and $W^2_{\pm}=M_V\sqrt{s_{NN}}e^{\pm |\mathrm{y}|}$.
 $S^2(W_\pm)$ is the survival factor that has been studied in \cite{Jones:2016icr}. $dn/dk$ is the photon flux of the proton.
We use the survival factors in LHCb proton-proton ultraperipheral collisions from \cite{Jones:2016icr}. At the NICA  energies, we use the survival factors equal to unity in accordance with \cite{Jones:2016icr}.

\begin{figure}[!h]
	\centering
	\includegraphics[width=6.7cm]{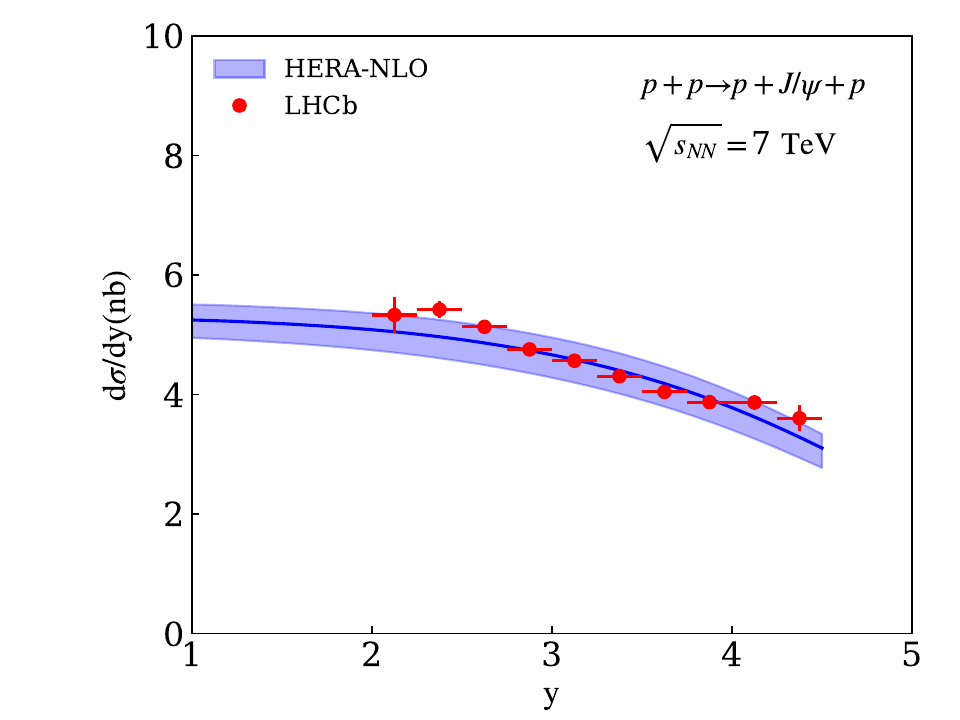}
	\includegraphics[width=6.7cm]{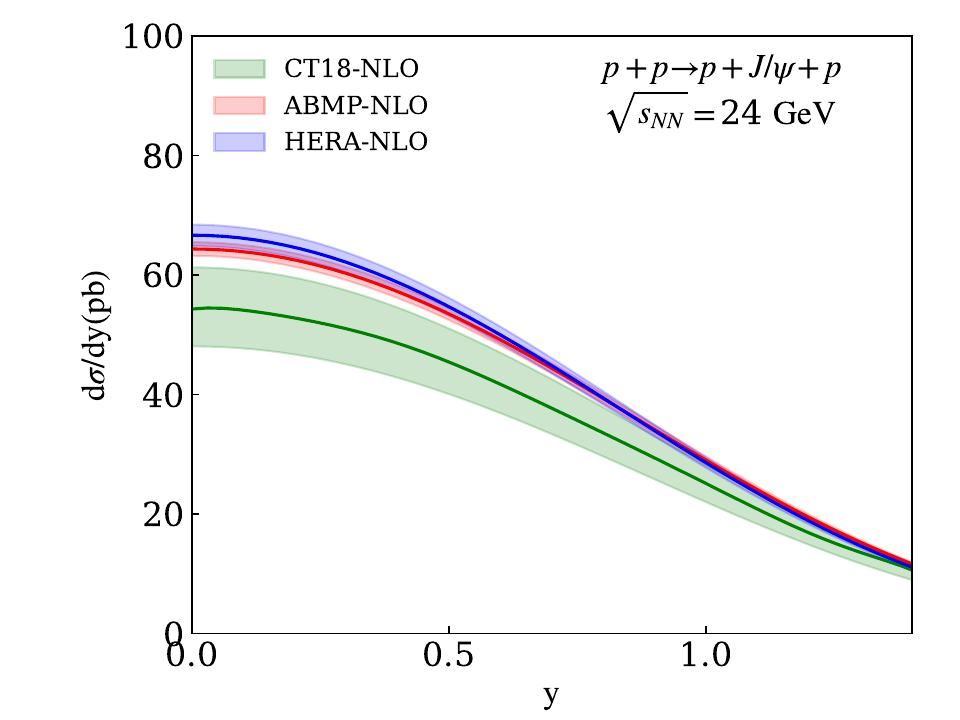}
	\caption{Left: Exclusive $J/\psi$ production in proton-proton ultraperipheral collisions as a function of rapidity at LHCb, and for NICA energy range, right graph. }
	\label{fig03}
\end{figure}

The exclusive $J/\psi$ production was measured at LHCb in proton-proton ultraperipheral collisions.
The model results for $J/\psi$ cross section in proton-proton ultraperipheral
сollisions  as a function of rapidity for GPDs based on the HERA gluon density are shown in Fig \ref{fig03}, left. Our results describe LHCb data well. The uncertainties in cross section are quite small because  the gluon density uncertainties at small-x are small.

We predict the $J/\psi$ cross section at NICA $\sqrt{s_{NN}}$ = 24 GeV. Our
results are presented in Fig. \ref{fig03}, right. The cross section at NICA is not very high because $J/\psi$ meson production here is not far from the threshold.
The model estimations for the $J/\psi$ cross section at $\sqrt{s_{NN}}$ = 24 GeV can help to simulate the detector system of NICA. Generally, the gluon density uncertainties are strongly correlated with $J/\psi$ production cross section uncertainties. This means that the vector meson production in $pp$ ultraperipheral collisions can be employed to obtain an additional
constrain for the uncertainties of gluon density.
\vspace{9mm}

To conclude, we can mention that we study the exclusive production of $J/\Psi$ meson within
a handback factorization approach. We analyze heavy meson production in lepton-proton and  ultraperipheral $pp$ collisions. The obtained cross sections of $J/\Psi$ production in a wide energy
range are in good agreement with the experiment. We perform prediction for spin asymmetries for the energy range of future colliders. Important information on  gluon GPDs (especially
 $E_g$ and $\tilde H_g$) can be obtained at US EIC, China  EicC and NICA.\\

 This work is partially supported by CAS president's international fellowship initiative (Grant No. 2025VMA0005),  National Key R\&D Program of China (Grant No. 2024YFE0109800) and the NFSC grant (Grant No. 12293061) .


\begin{thebibliography}{1}
\def\selectlanguageifdefined#1{
\expandafter\ifx\csname date#1\endcsname\relax
\else\selectlanguage{#1}\fi}
\providecommand*{\href}[2]{{\small #2}}
\providecommand*{\url}[1]{{\small #1}}
\providecommand*{\BibUrl}[1]{\url{#1}}
\providecommand{\BibAnnote}[1]{}
\providecommand*{\BibEmph}[1]{\emph{#1}}
\ProvideTextCommandDefault{\cyrdash}{\hbox to.8em{--\hss--}}
\providecommand*{\BibDash}{\ifdim\lastskip>0pt\unskip\nobreak\hskip.2em\fi
\cyrdash\hskip.2em\ignorespaces}

\bibitem{GPD-review01}
\selectlanguageifdefined{english}
\BibEmph{Goeke~K.,Polyakov~M.~V. and Vanderhaeghen~M.} {Hard exclusive reactions and the structure of hadrons}~//
 {Prog. Part. Nucl. Phys.}
  \BibDash
\newblock 2001. \BibDash
\newblock V.~47. \BibDash
\newblock P.~401--515. \BibDash
\newblock arXiv:hep-ph/0106012 [hep-ph].

\bibitem{GPD-review02}
\selectlanguageifdefined{english}
\BibEmph{Diehl~M.} {Generalized parton distributions}~//
  {Phys. Rept.}
  \BibDash
\newblock 2003. \BibDash
\newblock V.~388. \BibDash
\newblock P.~41--277. \BibDash
\newblock arXiv:hep-ph/0307382 [hep-ph].

\bibitem{burk}
\selectlanguageifdefined{english}
\BibEmph{Burkardt~M.} {Impact parameter space interpretation for generalized parton distributions}~//
  {Int. J. Mod. Phys. A}
  \BibDash
\newblock 2003. \BibDash
\newblock V.~18. \BibDash
\newblock P.~173--208. \BibDash
\newblock arXiv: hep-ph/0207047 [hep-ph].


\bibitem{dvcs}
\selectlanguageifdefined{english}
\BibEmph{Collins~J.~C., Frankfurt~L. and Strikman~M.} {Factorization for hard exclusive electroproduction of mesons in QCD}~//
  {Phys. Rev. D.}
  \BibDash
\newblock 1997. \BibDash
\newblock V.~56. \BibDash
\newblock P.~2982--3006. \BibDash
\newblock arXiv:hep-ph/9611433 [hep-ph].

\bibitem{dvmp2}
\selectlanguageifdefined{english}
\BibEmph{Vanderhaeghen M., Guichon P.~A.~M. and Guidal M.} {Deeply virtual electroproduction of photons and mesons on the nucleon: Leading order amplitudes and power corrections}~//
  {Phys. Rev. D}
  \BibDash
\newblock 1999. \BibDash
\newblock V.~60. \BibDash
\newblock P.~094017. \BibDash
\newblock arXiv:hep-ph/9905372 [hep-ph].

\bibitem{ji}
\selectlanguageifdefined{english}
\BibEmph{Ji X.} {Deeply virtual Compton scattering}~//
  {Phys. Rev. D.}
  \BibDash
\newblock 1997. \BibDash
\newblock V.~55. \BibDash
\newblock P.~ 7114--7125. \BibDash
\newblock arXiv: hep-ph/9609381 [hep-ph].

\bibitem{rad}
\selectlanguageifdefined{english}
\BibEmph{Radyushkin A.V.} {Asymmetric gluon distributions and hard diffractive electroproduction}~//
  { Phys. Lett. B.}
  \BibDash
\newblock 1996. \BibDash
\newblock V.~385. \BibDash
\newblock P.~333-342. \BibDash
\newblock arXiv:   hep-ph/9605431 [hep-ph].

\bibitem{rys1}
\selectlanguageifdefined{english}
\BibEmph{Ryskin M.~G.} {Diffractive $J/ \Psi$ electroproduction in LLA QCD}~//
  {Z. Phys. C.}
  \BibDash
\newblock 1993. \BibDash
\newblock V.~57. \BibDash
\newblock P.~89--92.

\bibitem{rys2}
\selectlanguageifdefined{english}
\BibEmph{Martin A.~D., Nockles C., Ryskin M.~G. and Teubner T.} {Small x gluon from exclusive $J/\Psi$ production}~//
  {Phys. Lett. B.}
  \BibDash
\newblock 2008. \BibDash
\newblock V.~662. \BibDash
\newblock P.252-258. \BibDash
\newblock arXiv:  0709.4406 [hep-ph].

\bibitem{jpsi}
\selectlanguageifdefined{english}
\BibEmph{Goloskokov S.~V., Xie Y.~P. and Chen X.} {Study of gluon GPDs in exclusive $J/\Psi$ production in electron-proton scattering}~//
  {Phys. Rev. D.}
  \BibDash
\newblock 2024. \BibDash
\newblock V.~110, no. 7. \BibDash
\newblock P.076029. \BibDash
\newblock arXiv: 2408.05800 [hep-ph].

\bibitem{eic}
\selectlanguageifdefined{english}
\BibEmph{Accardi A., Albacete J.~L.,Anselmino M.~et al.} {Electron Ion Collider: The Next QCD Frontier: Understanding the glue that binds us all}~//
  {Eur. Phys. J. A.}
  \BibDash
\newblock 2016. \BibDash
\newblock V.~52, no. 9. \BibDash
\newblock P. 268. \BibDash
\newblock arXiv: 1212.1701 [nucl-ex].


\bibitem{eicc}
\selectlanguageifdefined{english}
\BibEmph{Anderle D.~P., Bertone V., Cao X..~et al.} {Electron-ion collider in China}~//
  {Front. Phys. (Beijing).}
  \BibDash
\newblock 2021. \BibDash
\newblock V.~16, no. 6. \BibDash
\newblock P. 64701. \BibDash
\newblock arXiv:2102.09222 [nucl-ex].


\bibitem{jpsipp}
\selectlanguageifdefined{english}
\BibEmph{Xie Ya-Ping, Goloskokov S.~V.} {Exclusive $J/\Psi$ production in proton–proton collisions adopting GPD approach}~//
  {Eur.Phys. J. C.}
  \BibDash
\newblock 2025. \BibDash
\newblock V.~85, no. 6. \BibDash
\newblock P. 680. \BibDash
\newblock arXiv: 2502.17743 [hep-ph].

\bibitem{gk06}
\selectlanguageifdefined{english}
\BibEmph{Goloskokov S.~V. and Kroll P.} {Vector meson electroproduction at small Bjorken-x and generalized parton distributions}~//
  {Eur.Phys. J. C.}
  \BibDash
\newblock 2005. \BibDash
\newblock V.~42. \BibDash
\newblock P. 281--301. \BibDash
\newblock arXiv: hep-ph/0501242[hep-ph].

\bibitem{mus99}
\selectlanguageifdefined{english}
\BibEmph{Musatov I.~V. and Radyushkin A.~V.} {Evolution and models for skewed parton distributions}~//
  {Phys. Rev. D.}
  \BibDash
\newblock 2000. \BibDash
\newblock V.~61. \BibDash
\newblock P. 074027. \BibDash
\newblock arXiv: hep-ph/9905376 [hep-ph].


\bibitem{Koempel:2011rc}
\selectlanguageifdefined{english}
\BibEmph{Koempel J., Kroll P.,Metz A. and Zhou J.} {Exclusive production of quarkonia as a probe of the generalized parton distribution for gluons}~//
  {Phys. Rev. D.}
  \BibDash
\newblock 2012. \BibDash
\newblock V.~85. \BibDash
\newblock P. 051502. \BibDash
\newblock arXiv: 1112.1334 [hep-ph].

\bibitem{gk08}
\selectlanguageifdefined{english}
\BibEmph{Goloskokov S.~V. and Kroll P.} {The Target asymmetry in hard vector-meson electroproduction and parton angular momenta}~//
  {Eur.Phys. J. C.}
  \BibDash
\newblock 2009. \BibDash
\newblock V.~59. \BibDash
\newblock P. 809-819. \BibDash
\newblock arXiv: 0809.4126 [hep-ph].

\bibitem{Jones:2016icr}
\selectlanguageifdefined{english}
\BibEmph{Jones S.~P., Martin A.~D, Ryskin M.~G. and Teubner T.} {Exclusive $J/\psi$ production at the LHC in the $k_T$ factorization approach}~//
  {J. Phys. G}
  \BibDash
\newblock 2017. \BibDash
\newblock V.~44, no. 3. \BibDash
\newblock P. 03LT01. \BibDash
\newblock arXiv: 1611.03711 [hep-ph].



\end{thebibliography}
\end{document}